# Enhanced Inter-cell Interference Coordination for Heterogeneous Networks in LTE-Advanced: A Survey

Lars Lindbom (Ericsson), Robert Love, Sandeep Krishnamurthy (Motorola Mobility), Chunhai Yao (Nokia Siemens Networks), Nobuhiko Miki (NTT DOCOMO), Vikram Chandrasekhar (Texas Instruments)

December 7, 2011
Contact: vikram.chandrasekhar@ti.com

## Abstract

Heterogeneous networks (het-nets) – comprising of conventional macrocell base stations overlaid with femtocells, picocells and wireless relays – offer cellular operators burgeoning traffic demands through *cell-splitting gains* obtained by bringing users closer to their access points. However, the often random and unplanned location of these access points can cause severe near-far problems, typically solved by coordinating base-station transmissions to minimize interference. Towards this direction, the 3$^{rd}$ generation partnership project Long Term Evolution-Advanced (3GPP-LTE or Rel-10) standard introduces time-domain inter-cell interference coordination (ICIC) for facilitating a seamless deployment of a het-net overlay. This article surveys the key features encompassing the physical layer, network layer and back-hauling aspects of time-domain ICIC in Rel-10.

## 1  Introduction

Global mobile data traffic has tripled each year since 2008 and it is projected to increase a whopping 26-fold between 2010 and 2015. By 2015, 4 major regions (Sub-Saharan Africa, Southeast Asia, South Asia, and the Middle East) and 40 countries (including India, Indonesia, and Nigeria) will have more people with mobile network access than with access to electricity at home [1]. To address this exponential growth in demand, cellular operators are aggressively overlaying smaller form factor low power nodes (also known as *small cells)* on top of existing macrocell base stations. Being an infrastructure overhaul to existing macrocell eNodeBs,[1] het-nets help offload users on to small cells, thereby enabling higher quality of service to users with otherwise poor coverage. Because of the smaller transmission powers from small cells and the relative proximity of users to a small cell, more users can now be packed within the same area in a het-net; consequently, het-nets provide *"cell-splitting"* gains relative to macro-only networks.

---

[1] The LTE radio access network architecture is designed as a *flat architecture* consisting of just one type of node, known in LTE as an enhanced NodeB (eNodeB). The acronym UE stands for the user equipment (mobile terminal).

With het-net deployment in same spectrum, users can experience severe interference (also called the loud neighbor effect). The loud neighbor effect is due to the often geographically random low power node deployment as well as the near-far problem arising from the imbalance in path-gains and transmission powers between the macrocell and low power nodes [2]. 3GPP-LTE has devoted significant standardization effort towards devising inter-cell interference coordination (ICIC) schemes for minimizing interference, culminating in the so-called "enhanced" ICIC in LTE-Advanced.

This article surveys the key features encompassing the physical layer, network layer and back-hauling aspects of ICIC in LTE, with specific focus on Rel-10 enhanced ICIC.

## 2 The LTE Cell Architecture

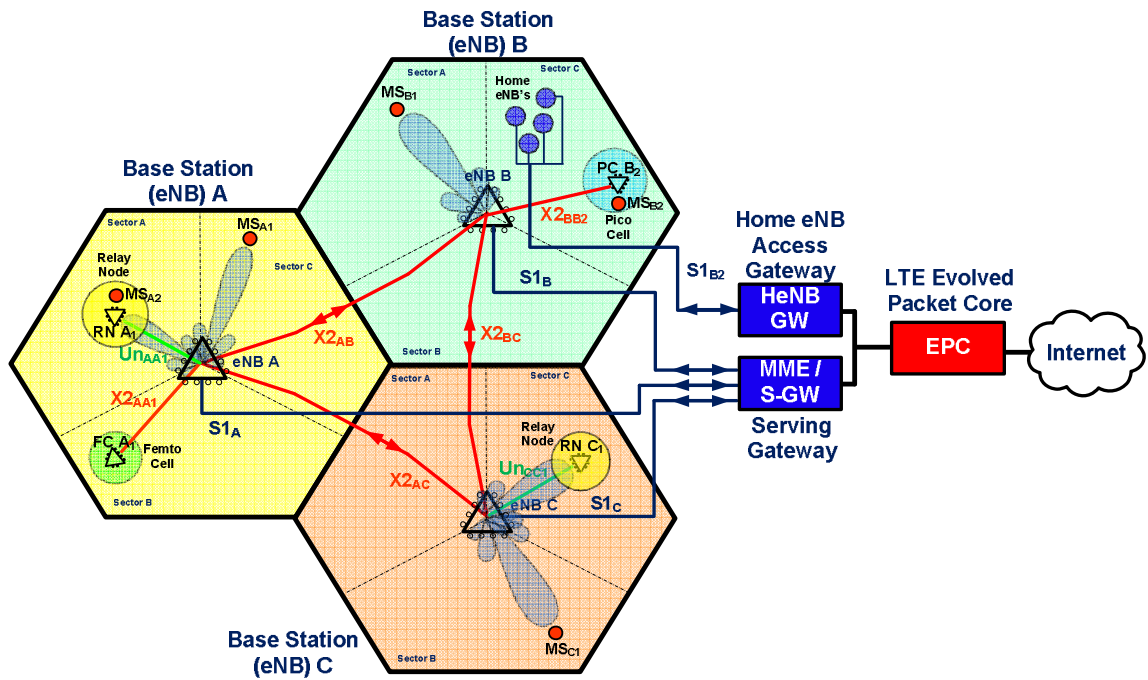

**Figure 1 – Access via LTE Macrocell nodes, Picocell nodes, Relay nodes, and Femtocell and HeNB nodes**

Figure 1 shows a generic LTE architecture. The access network of LTE, called E-UTRAN, consists of a network of eNodeBs connected via different interfaces. In LTE, eNodeBs are normally inter-connected with each other by means of an interface called X2 and to the core network through an interface called S1. Although 3GPP-LTE employs a flat architecture, for purposes of exposition, cells can be classified in terms of their transmission powers, antenna heights, the type of access mechanism provided to users, and the backhaul connection to other cells.

**Macrocells** cover a large cell area (typical cell radius being of the order of 500 meters to a kilometer), with transmit antennas above the clutter and transmission power of the order of 46 dBm (20 watts). They provide service to all users.

**Femtocells,** also called **Home eNodeBs (HeNBs)** are lower power cells installed (typically indoors) by the end-consumer. HeNB access is classified either as a closed, hybrid or open access type. A closed access HeNB maintains a Closed Subscriber Group (CSG) white-list where access is limited only to subscribed users (i.e. to UEs that are members of the CSG). Hybrid access HeNBs allow limited access to non-subscribed UEs, but provide differentiated higher quality of service to CSG users. Open access HeNBs provide undifferentiated access to all UEs. In Rel-10, X2 interface is used between open access HeNBs and between closed/hybrid access HeNBs with identical CSG IDs and between closed/hybrid HeNBs and open access HeNBs.

**Picocells** are operator deployed cells, with lower transmission powers – typically an order of magnitude smaller – relative to macrocell eNodeBs. They are installed typically in wireless hotspot areas (for example, malls) and provide access to all users.

**Relay Nodes** are operator deployed and are primarily used to improve coverage in new areas (e.g. events, exhibitions etc.). Unlike HeNBs and picocells which connect to the macrocell over X2 backhaul, Relay nodes backhaul their traffic through a wireless link to a *Donor* eNodeB. Inband relays use the same frequency of operation over their backhaul link as the access (relay-UE) links. Outband relays, on the other hand, use different spectrum over backhaul and access link.

# 3   LTE Downlink Slot Structure

In 3GPP-LTE, downlink transmissions employ orthogonal frequency division multiple access (OFDMA) transmission. The basic unit of transmission is a downlink subframe, defined by two slots, each having duration of 0.5 milli-seconds. The transmitted signal (Figure 2) over each antenna in each slot is described by N (lying between 6 and 110) physical resource blocks [2] (PRBs) which defines the mapping of certain physical channels to resource elements.

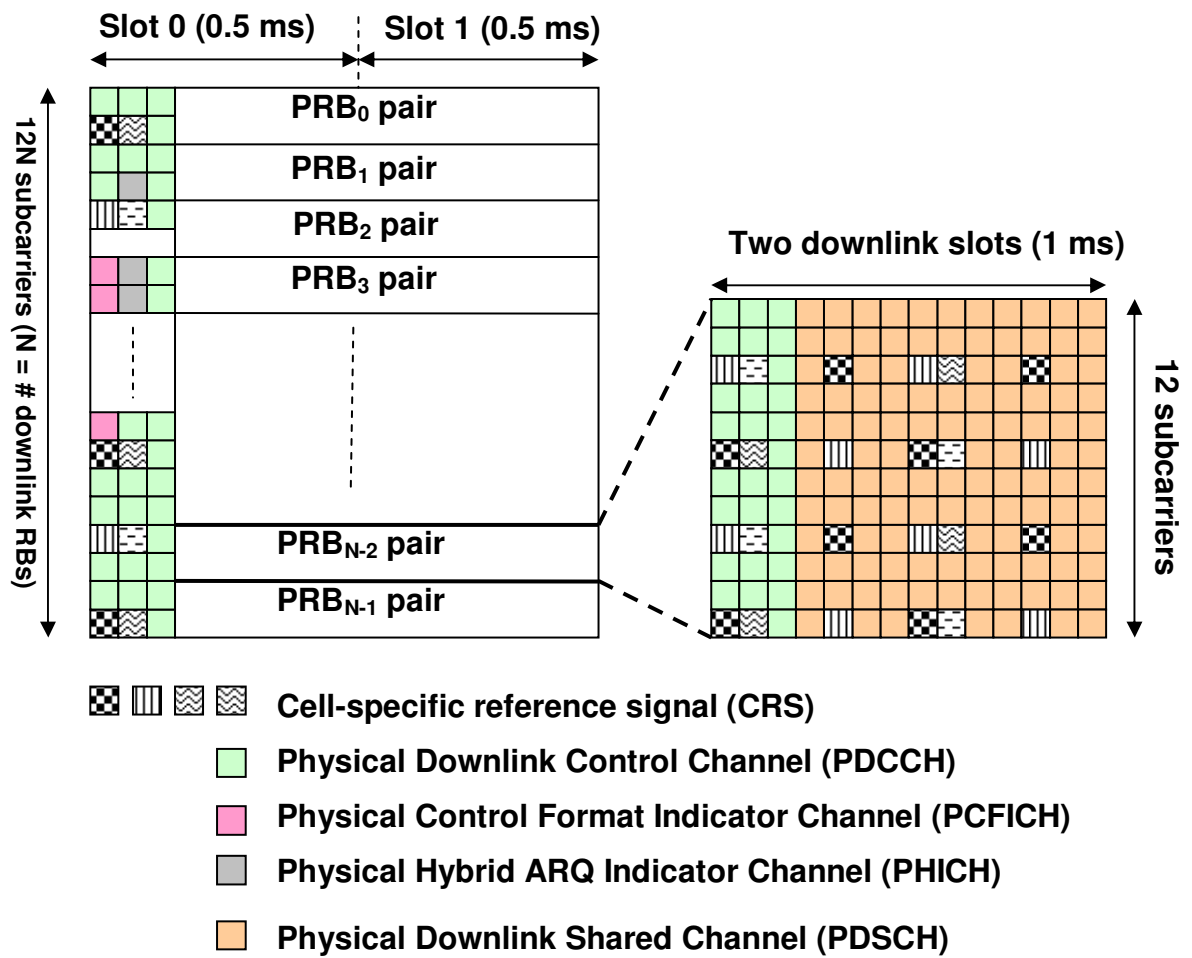

Figure 2: Physical Slot Channel Structure

---

[2] Each physical resource block (PRB) occupies one slot (0.5 ms) in time and 180 KHz (12 subcarriers) in frequency.

A downlink subframe therefore comprises of N concatenated PRB pairs. Cell-Specific Reference Signals (CRS) are pilot signals, essential for downlink demodulation, mobility measurements and to acquire channel-state information (CSI). The control payload for all users is carried via the Physical Downlink Control Channel (PDCCH) whose duration is conveyed via the Physical Control Format Indicator Channel (PCFICH). The control payload conveys the RB assignment for the uplink information data (Physical Uplink Shared Channel or PUSCH) and downlink information data (Physical Downlink Shared Channel or PDSCH), while hybrid automatic-repeat-request acknowledgments for PUSCH are carried by the PHICH signal. While the PDCCH is transmitted across the entire bandwidth in the first slot of each subframe, the PDSCH signal is transmitted – on specific PRB locations signaled via PDCCH – to scheduled user(s).

## 4  Rel-8/9 ICIC: Frequency Domain ICIC

In Rel-8, ICIC is performed over the frequency domain by transmission of messages across a standardized backhaul interface such as X2. Because the latency over X2 is typically of the order of a tens of milli-seconds (or a few radio frames), it is expected that any updates (reconfiguration) to the ICIC messages are relatively infrequent (order of seconds).

Frequency-domain ICIC over downlink in Rel-8 is based on an event triggered message exchanged between eNodeBs as often as every 200 ms to allow, for example, adaptive fractional frequency reuse (FFR). Uplink ICIC is based on event triggered Overload Indicator (OI) and High Interference Indicator (HII) messages exchanged between eNodeBs as often as every 20 ms.

**Relative Narrow-Band Transmit Power (RNTP):** The RNTP message is signaled using a bitmap wherein each RNTP bit value indicates whether the corresponding PRB pair is limited by a transmit power threshold or not. Upon receipt of the RNTP message, the recipient eNodeBs can take into account this information while determining their scheduling decisions for subsequent subframes. One such example may be that the recipient eNodeB avoids scheduling UEs in PRBs where the source eNodeB (e.g., macrocell eNodeB) is transmitting above a certain power limit.

**High Interference Indicator (HII):** The HII is a proactive indicator sent as a bitmap with one bit per PRB/subband that the serving cell intends to use for scheduling cell-edge UEs potentially causing high Inter-Cell Interference. An eNodeB can send HII with different, neighbor-cell specific contents to different neighbor cells.

**Overload Indicator (OI):** The OI is a reactive indicator exchanged over the X2 reflecting the uplink interference plus noise level (low, medium, or high) of a PRB measured by an eNodeB. Although the reaction of the recipient eNodeB upon receipt of the OI is not standardized, one OI method is for the recipient eNodeB to limit the maximum transmit power of a UE scheduled with PRBs indicated by OI to have high interference

In the following sections, we discuss different het-net deployment scenarios and motivate the need for enhanced ICIC in these scenarios.

## 5 Deployment scenarios motivating enhanced ICIC

### 5.1 Motivation for ICIC

Two different het-net scenarios were investigated during the 3GPP-LTE Rel-10 work item phase: (1) co-channel macro-femto deployment and (2) co-channel macro-pico deployment. In macro-femto deployments, macrocell UEs may experience large interference when they move close to CSG-HeNBs. In macro-pico deployments, interference may be larger at picocell UEs as a result of increasing the picocell radio range in order to offload greater numbers of users from the macrocell eNodeB on to picocells.

### 5.2 CSG Femtocell Deployment Aspects

In the co-channel macro-femto scenario, a macrocell UE must remain attached to its serving macrocell when it is in close proximity of a CSG femtocell if it does not belong to the femtocell's CSG group, resulting in unreliable PDCCH reception (see Section 5.4).

During 3GPP studies it was found that a macro UE will, on an average, experience a PDCCH coverage hole, 20% of the time [3]. Because the macro UE cannot decode its PDCCH reliably, it cannot be reliably scheduled over the downlink. Making matters worse, the UE will statistically experience a coverage hole for the physical broadcast channel (PBCH) – containing important system information for initial acquisition – roughly 15% of the time [3].

This motivated the need for an ICIC scheme that could enable the CSG HeNBs to reduce their downlink interference by either reducing their transmission powers or avoiding scheduling their users on some resources so that macro UEs could be protected from CSG transmissions.

## 5.3 Picocells Deployment Aspects

In the macro-pico scenario, two problems that are caused by the difference in the transmission power between the macrocell eNodeB and picocell eNodeB are shown in Fig. 3.

**Downlink/Uplink Imbalance**. The downlink coverage of the macrocell eNodeB is much larger than that of the picocell eNodeBs. On the other hand, the difference in transmission power does not affect the coverage in the uplink, since the transmitter is the UE. Therefore, the eNodeB that provides the best downlink coverage may be different from the eNodeB providing best uplink coverage.

**Cell Range Expansion.** The second problem is that the number of UEs connected to the picocell eNodeBs is much smaller than that of macrocell eNodeBs resulting in inefficient resource utilization. It is beneficial for the network to bias handover preferentially towards the picocell eNodeBs, e.g., add a handover offset to the picocell eNodeB reference signal received signal power (RSRP) so that the UE preferentially selects a picocell eNodeB even when it is not the strongest cell. This method is called **Cell Range Expansion (CRE)**.

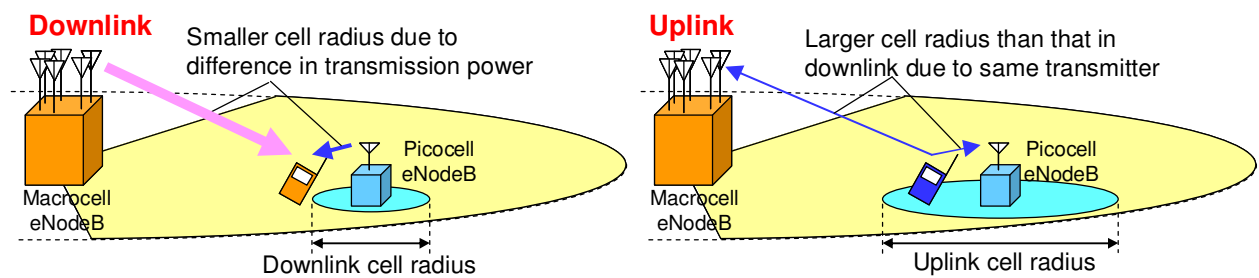

(a) Uplink downlink imbalance.

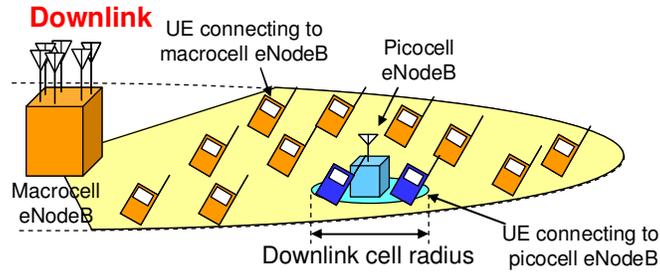

(b) Smaller number of UEs connected to picocell eNodeB.

Figure 3. Problems in macro-pico scenario.

Although CRE enables higher user offloading from macrocell eNodeBs on to picocells, different problems can arise due to the fact that the UE's serving picocell eNodeB is not its strongest cell. The UEs connecting to the picocell eNodeB with large-bias CRE can suffer from severe interference from the aggressor macrocell eNodeB since the received signal power of the macrocell eNodeB is larger than that of picocell eNodeB for such UEs.

## 5.4 Motivating Rel-10 ICIC Enhancements

One may justifiably ask why Rel-8 ICIC – PRB coordination across different eNodeBs – is not sufficient to address the aforementioned problems in both macro-femto and macro-pico scenario. In Rel-8/9, the PDSCH can be transmitted within a limited bandwidth by using the PRB-level multiplexing in order to achieve frequency-domain scheduling gains. This design, alongside the RNTP mechanism in Rel-8 ICIC therefore enables frequency-domain ICIC for PDSCH transmission in single-carrier operation.

On the other hand, the Rel-8 control payload (carried via the PDCCH) is transmitted over the entire bandwidth to achieve frequency domain diversity gain. Therefore, the Rel-8 PDCCH signal cannot take advantage of RB-level coordination and scheduling since the PDCCH signals are scrambled and interleaved across the first several symbols of a subframe. Each UE must monitor for its potential PDCCH by performing blind decoding at PDCCH candidate locations for different PDCCH types.

Interference problems for PDCCH and other control channels in the LTE DL control region are typically avoided via interference randomization from interleaving and power de-boosting.

Interference problems can become significant when large bias CRE (> 6 dB) is applied to keep UEs attached to, for example, a picocell or when macrocell-UEs must remain attached to their serving macrocell when they are in close proximity of a non-allowed CSG femtocell.

Since the control payload for a given UE is potentially distributed across the entire system bandwidth, a UE may experience high levels on interference on its PDCCH signal even though interference on its PDSCH signal can be overcome using Rel-8 ICIC. As a result, Rel-10 ICIC studies have focused on improving signal reception at UEs that would otherwise experience coverage holes on their control channel signals.

# 6 Overview of Rel-10 ICIC

During Rel-10, 3GPP developed two approaches to coordinate inter-layer interference, primarily targeting the aforementioned deployment scenarios. Both approaches rely on a semi-static resource partitioning across layers but explores two different dimensions for ICIC of control and data channels. In one approach, interference avoidance is addressed by means of ICIC in frequency domain, relying on carrier-aggregation as introduced in Rel-10. In the other approach, interference avoidance is addressed by means of ICIC in time domain, relying on so-called almost blank subframes. Both approaches target scenarios with low power nodes being time synchronized to the macrocell eNodeB.

## 6.1 Carrier-aggregation based ICIC

In carrier-aggregation (CA), a terminal may receive and/or transmit on multiple component carriers simultaneously, where a component carrier is a fully backward compatible LTE carrier. A terminal supporting carrier aggregation can be configured by higher layer signaling to enable cross-carrier scheduling on certain component carriers. This implies e.g. that a terminal receiving a downlink assignment on one component carrier may receive associated data on another component carrier. One of the main motivations for introducing cross-carrier scheduling was to enhance operations in heterogeneous networks in a multi-carrier deployment.

The basic idea with CA-based ICIC is to create a "protected" component carrier for reliable reception of downlink physical signals, system information and control channels at victim layers but where data can be received on any configured downlink component carrier via cross-carrier scheduling. This, in conjunction with RNTP at aggressor layers (see section 2), forms the basis of CA-based ICIC.

Consider a macro-pico deployment scenario with two DL component carriers, CC1 and CC2. In this particular scenario, a Macro creates a protected component carrier by reducing its transmit power on CC1 in order to enable larger picocell coverage on CC1. On CC2, the macrocell transmits with its nominal output power implying that the size of the picocell associated with CC2 will be reduced in comparison with the picocell coverage on CC1. This scenario is illustrated by Figure 4a) where dashed areas represent so-called cell range expansion (CRE) zones, i.e. areas with reduced reception reliability of e.g. the downlink control channels PDCCH, PHICH and PCFICH. From this figure, it is clear that without ICIC, a terminal configured with CC1 and CC2 can simultaneously receive on both carriers only in the following scenarios:

- it operates within the RSRP coverage of the picocell associated with CC2 when connected to the picocell.

- it operates within the RSRP coverage of the macrocell on CC1 when connected to the macrocell.

However, with cross-carrier scheduling a picocell edge user can receive PDCCH on the protected CC1 whereas the associated PDSCH are received on CC2, on which the Macro will have reduced their transmit power on PDSCH in resource blocks known by the Pico via X2 signaling of RNTP. Figure 4b) illustrates the operation of the CA-based ICIC approach in which the Macro avoids transmissions (or reduce its transmission power) of PDSCH in resource blocks in the upper bandwidth of CC2. In this illustration, users operating in the cell range expansion zones receive data via cross-carrier scheduling whereas users closer to the eNodeB sites do not need to rely on data reception via cross-carrier scheduling.

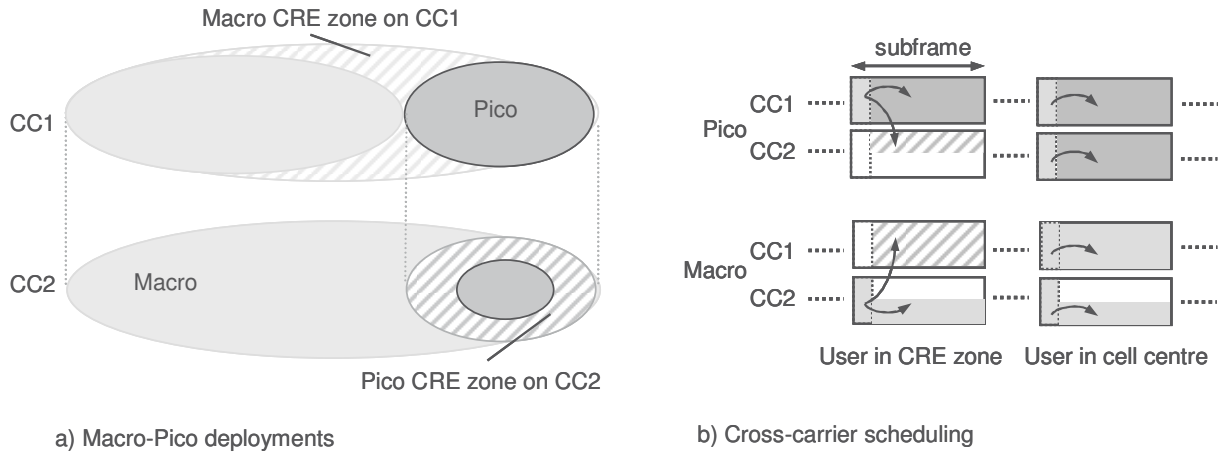

Figure 4. Carrier-aggregation based ICIC

In CA-based ICIC, a primary cell associates with a "protected" component carrier, so in Figure 4. Picocell users would then have CC1 as their primary cell whereas macrocell users would have CC2 as their primary cell.

CRS transmission from an aggressor cell (necessary for backwards-compatibility) will potentially create inter-cell interference towards victim users Rather than rely on CRS for data demodulation and CSI reporting, Rel-10 introduced a new transmission mode (TM9) for which demodulation of PDSCH relies entirely on pilot signals called the UE-specific reference signals (UE-RS) for data demodulation and CSI reference signals (CSI-RS) for channel quality estimation. Through RNTP signaling, UE-RS collisions across different cells can be avoided since UE-RS are only transmitted in RBs carrying PDSCH. Such "UE-specific" configuration is therefore useful in het-net scenarios where CRS reception is potentially deteriorated. For example, a picocell user operating in the CRE zone of CC2 in Figure 4a) would likely be configured with TM9 on CC2 as the received signal strength of the corresponding CRS could be weak in the CRE zone. Furthermore, a terminal can be configured to base measurements for mobility and radio link monitoring on its primary cell only, i.e. the cell with good CRS reception.

## 6.2 Time-domain based ICIC

The basic idea with time domain ICIC is that an aggressor layer creates "protected" subframes for a victim layer by reducing its transmission activity in certain subframes. To do so, the aggressor eNodeB reduces its transmission power of some downlink signals (or alternatively, mutes their transmission) during a set of low interference subframes designated as **Almost-Blank Subframes (ABS)** whose occurrences are known *a priori* at the coordinating eNodeBs (see Figure 5). ABS subframe patterns can be constructed by configuring so-called multicast/broadcast over single-frequency network (MBSFN) subframes or/and by not scheduling unicast traffic (or by reducing transmit powers) in certain subframes.

During ABS subframes, the aggressor eNodeB does not transmit PDSCH but, may transmit CRS, critical control channels, and broadcast and paging information (for ensuring legacy device support). If the victim cell schedules its UEs in subframes that overlap with aggressor cell ABS transmissions, the victim cell "protects" its UEs from strong inter-cell interference, thereby improving the chances of successful PDCCH reception.

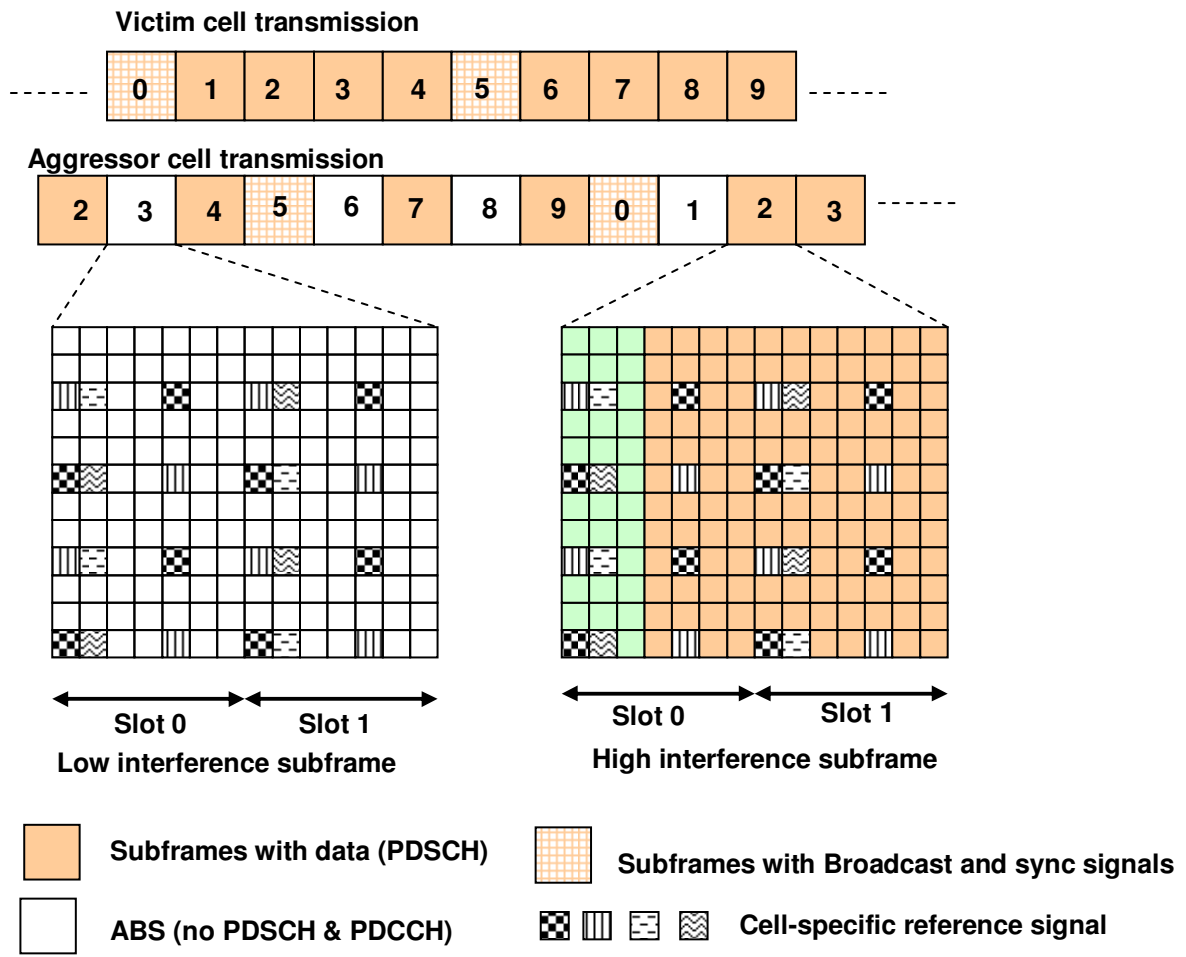

Figure 5: Time-domain inter-cell coordination in LTE-Advancced

Backward compatibility is ensured by transmitting physical signals, system information and paging in the same way as in Rel-8. This implies that CRS transmission at the aggressor layers will create interference on either PDSCH signal or CRS signals at victim layers even during ABS subframes, although the CRS interference on PDSCH can be reduced by configuring MBSFN subframes, since the CRS is not transmitted in PDSCH region in MBSFN subframes. Furthermore, PSS/SSS and system information are transmitted in subframe 0 and 5 of a radio frame, and will thus be inter-layer interference sources that cannot be avoided by introducing ABS.

The operation of time-domain based ICIC in a practical deployment is illustrated in Figure 6. In the case of a macro-femto deployment, CSG-femtocells represent the aggressor layer, whereas in the case of a macro-pico deployment, macrocells represent the aggressor layer. With time-domain ICIC, a macrocell eNodeB may schedule users during protected subframes when they are in the proximity of CSG-femtocells, whilst a picocell eNodeB may schedule their cell-edge users only during protected subframes.

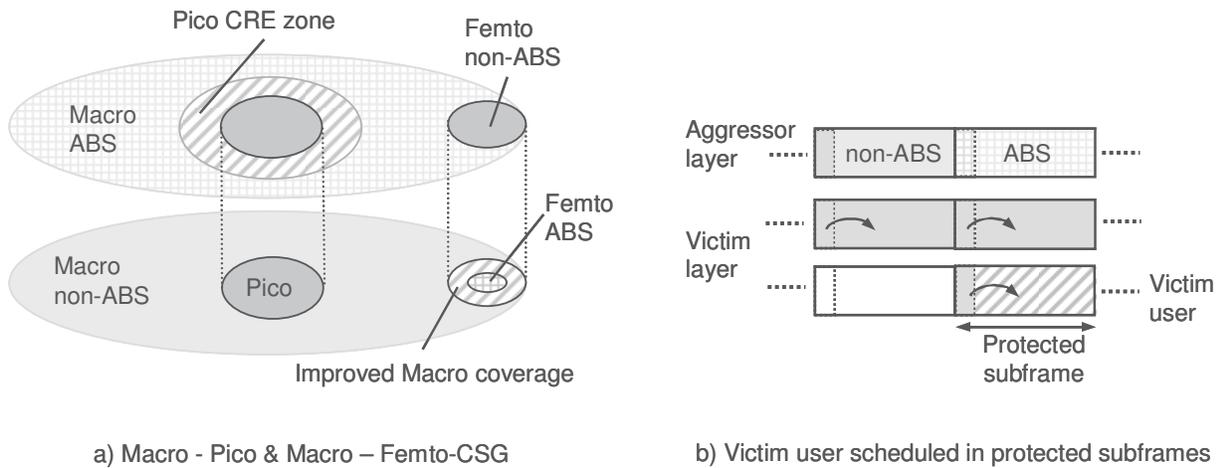

a) Macro - Pico & Macro – Femto-CSG        b) Victim user scheduled in protected subframes

Figure 6 Time domain based ICIC

The information regarding the subframes on which coordinated low interference subframe transmission occurs is exchanged between coordinating eNodeBs via bitmap patterns. Upon receiving the bitmap pattern – either via customized operation administration and maintenance (OAM) or via X2-interface – the recipient eNodeB (e.g., picocell eNodeB) can schedule data for its victim UEs on subframes that overlap with the aggressor ABS transmissions. In order to enable this, time-domain ICIC requires time-synchronized eNodeB transmissions at least at the subframe boundaries.

For optimal operation with synchronous uplink Hybrid-ARQ (with 8 ms round-trip-time) in a FDD het-net, a bitmap of length 40 bits (which is equivalent to a 40 ms duty cycle periodicity) is used. Also, as shown in Figure 5, additional interference mitigation through subframe shifting (possible only in LTE-FDD) is required for avoiding cross-layer interference from broadcast and synchronization channel signals (transmitted on subframes 0 and 5 in each radio frame). It is also desirable to configure paging occasions, and scheduled system information, in non-ABS subframes at the aggressor layer and in protected subframes at the victim layer.

## 7  Rel-10 ICIC: Resource restricted CSI measurements

In Rel-8, a UE may average channel and interference estimates over multiple subframes to derive a CSI feedback. When time-domain ICIC is employed, the interference level of these two types of subframes, i.e., protected and non-protected subframes can be significantly different. In order that the UE does not average interference across the two different types of subframes, the eNodeB can configure two subframe subsets such that the UE only averages the channel and interference estimates over subframes within the subset but, not across the two subsets. The UE then reports the CSI measurements separately for the two subsets either periodically (with the respective configured reporting periods for each subset) or aperiodically (i.e., one of the CSI measurements from two subframe subsets is reported via PUSCH when triggered by PDCCH).

## 8  Rel-10 ICIC: Handover Coordination and Mobility

In connected mode operation (where UE's mobility is <u>controlled by the network</u>), the network may perform an inter-frequency or inter-Radio Access Technology (inter-RAT) handover (HO) to a different carrier frequency when the network determines that a non-CSG UE [range extended picocell UE] is in the proximity of a CSG cell [resp. a macrocell eNodeB] based on the UE's measurement reports. In idle mode (where UE's are <u>responsible for their mobility within the network</u>), the UE autonomously performs an inter-frequency or inter-RAT reselection to an

eNodeB on a different carrier when the UE roams close to a dominant eNodeB interferer. Thus, in both idle mode and connected mode, an inter-frequency/inter-RAT HO can potentially create LTE coverage holes potentially leading to LTE service discontinuity in a co-deployed carrier. To avoid occurrence of LTE coverage holes, Rel-10 ICIC developed mobility support and radio link monitoring as discussed below.

## 8.1 Radio Resource Management (RRM)

In 3GPP-LTE, UE's perform RRM measurements for managing their connection and facilitating handover. In connected mode, the UE reports to the serving eNodeB one or both of the RSRP and the reference signal received quality (RSRQ) – an estimate of the ratio of the RSRP to the estimated total received power (signal plus interference and thermal noise). The network performs handover of a UE to a neighbor cell, for example, when the serving cell RSRP falls below the neighbor cell RSRP so as to ensure that the UE remains connected to the strongest cell even while mobile. Idle mode UEs, on the other hand, autonomously choose their best cell.

In Rel-10, a victim UE can be configured to perform resource-specific RRM measurements (e.g. only on protected subframes). Resource-specific RRM measurements can improve accuracy of RSRP/RSRQ reporting by eliminating persistent interference on the CRS ("collision" of CRS transmitted across different cells can be pre-emptively avoided via physical cell-identity planning).

## 8.2 Radio Link Monitoring (RLM)

RLM is a measurement function for ensuring that UEs maintain time-synchronization and can receive their control information reliably. In 3GPP-LTE, UEs evaluate the quality of the serving cell eNodeB downlink continuously and ensure that they are time-synchronized to the serving cell eNodeB. 3GPP-LTE is based on the principle of "*uplink after downlink*" in the sense that UEs should not transmit on uplink resources in the absence of a reliable downlink channel. Therefore, if downlink channel synchronization is lost, UEs must suspend their uplink transmissions and should either (a) try to recover their existing link or alternatively (b) try to find a new serving cell with which a reliable downlink channel can be established. RLM monitoring comprises **out-of-synchronization (OOS)** and **in-synchronization (IS)** evaluation procedures where UE monitors CRS quality of the serving cell in the presence of co-channel interference. OOS and IS are defined as events that the block error rates of hypothetical PDCCH transmissions corresponding to a pre-determined control payloads exceed pre-specified thresholds.

If the UE's serving cell is not the strongest cell, the magnitude and variance of its interference is potentially large. Such a UE can potentially declare a radio link failure (leading to loss in connection) if its detected OOS events are persistent. In order to ensure predictable UE behavior, Rel-10 provides a mechanism so that the serving eNodeB can configure *resource-specific RLM measurements* at the UE for capturing its radio link quality as experienced during protected subframes. This ensures that the link quality is tied to that of the protected subframes and the UEs can continue to maintain their connection.

## 9  Legacy Impact

Both the frequency-domain based Rel-8 ICIC and time-domain based Rel-10 ICIC schemes enable the victim cell to schedule its UEs over time-frequency resources that are lightly interfered with. It must be noted that Rel-8/9 UEs will not be able to utilize the newly introduced Rel-10 enhancements such as resource-restricted CSI measurements as well as resource restricted RLM/RRM measurements. Because interference levels can significantly vary between protected and non-protected subframes, it is likely that UE-autonomous functions such as RLM and idle mode RRM may be inadequate to sustain reasonable quality of service for Rel-8 terminals under high interference, although this problem can be avoided by applying CRE only to Rel-10 terminals.

## 10  Coexistence aspects of Rel-8/9 ICIC with Rel-10 ICIC

The specification does not mandate whether or not coordinating eNodeBs must employ a combination of Rel-8 ICIC (e.g. RNTP indicator) *and* Rel-10 ICIC in order to minimize mutual interference. One can immediately see that this could lead to a pathological situation wherein an eNodeB makes a "promise" to transmit with low power (e.g. using RNTP indicator) on certain PRBs even during subframes (e.g. during ABS subframes of a coordinating eNodeB) where it may be beneficial to schedule its users with full power across all PRBs.

Note however, that LTE, by design, is a "flat" architecture where there is no explicit hierarchy between different types of cells. Consequently, the behavior of an eNodeB in response to ICIC messages is left to implementation i.e. there is no explicit requirement that an eNodeB *is mandated* to coordinate its transmissions according to either its transmitted ICIC messages, or the ICIC messages received from coordinating cells.

## Biographies

**Robert Love** (S'81-M'84) graduated from the University of Florida, Gainesville in 1982 BSEE and 1984 MSEE after attending USAFA. From 1984 to 1986, he worked at Texas Instruments on communication and radar systems, and from 1986 to 1990, he worked at E-systems on signal processing and communication systems. Robert is currently a Fellow of the Technical Staff at Motorola Mobility and since 1990 has worked at Motorola (Mobility) in the wireless infrastructure and handset groups on products and research related to wireless communication systems such as GSM, CDMA, WCDMA, 1xEVDO/V, HSPA, 802.16e, LTE, LTE Advanced and 802.11. He has numerous publications and most recently contributed to the book "*LTE - The UMTS Long Term Evolution - From Theory to Practice*". His research interests include wireless communications, multimedia systems, and mobile applications and services. He has also been contributing to 3GPP standards since 1999 and is currently editor for one of the 3GPP-LTE physical layer specifications (TS 36.213).

**Sandeep Krishnamurthy** received his PhD from North Carolina State University in 2005 and the B. Tech. degree from Indian Institute of Technology – Madras in 2001, both in Electrical Engineering. He has been with Motorola since October 2005 and is currently a principal engineer. He has worked in GERAN towards GSM/GPRS/EGPRS evolution and in the 3GPP working groups towards LTE Rel-8/9/10 standardization. He has also been working as an algorithm designer and has participated in wireless modem prototyping for GPRS/EGPRS and LTE technologies within Motorola. His research interests are in areas of Signal Processing, Heterogeneous networks, Location technologies and RF engineering.

**Chunhai Yao** graduated from Harbin Institute of Technology and obtained Electronic Engineering bachelor degree in 1996 and master degree in 2001. He worked in Datang mobile as standard engineer from 2001 to 2004**,** focused on TD-SCDMA RF standard and system co-existence study. He joined in Siemens (now Nokia Siemens Networks) since 2004 as senior specialist of radio technology, participated LTE and LTE-Advanced standardization work in 3GPP RAN1; working area included heterogeneous networks and TD-LTE.

**Nobuhiko Miki** received his B.E. and M.E. degrees from Kyoto University, Kyoto, Japan in 1996 and 1998, respectively, and received his Dr. Eng. degree from Keio University, Yokohama, Japan in 2009. In 1998, he joined NTT Mobile Communications Network, Inc. (now NTT DOCOMO, INC.) His research interests include mobile communication systems.

**Vikram Chandrasekhar** is working as a design, systems and standards engineer in the communications infrastructure group at Texas Instruments. He works on various standardization aspects for 3GPP-LTE with emphasis on heterogeneous networks, relays and signalling enhancements. Prior to joining Texas Instruments, he received his PhD from the University of Texas under the supervision of Prof. Jeff Andrews in May 2009. His PhD dissertation focused on fundamental limits and algorithms for femtocell, picocell and hotspot-aided broadband cellular networks. He completed his B.Tech (honors) at Indian Institute of Technology, Kharagpur in 2000 and received his M.S. at Rice University in 2003. He is a recipient of the overall best paper award in the *IEEE Global Telecommunications Conference* held in Hawaii in 2009. He has also won the best communications paper in the *IEEE Asilomar Conference on Signals, Systems and Computers* held in 2008. He is a co-chair for the heterogeneous workshops held during the IEEE International Conference on Telecommunications (2011) and the IEEE Globecom (2011). He was a recipient of the National Talent Search scholarship awarded by the Government of India in 1994. He has held prior positions as a staff engineer in the RF high frequency measurement group at National Instruments and summer internships in Freescale and Texas Instruments.